# UPGRADE OF SPRING-8 BEAMLINE NETWORK WITH VLAN TECHNOLOGY OVER GIGABIT ETHERNET

M. Ishii, T. Fukui, Y. Furukawa, T. Nakatani, T. Ohata, R. Tanaka
SPring-8, Hyogo 679-5198, Japan

*Abstract*

The beamline network system at SPring-8 consists of three LANs; a BL-LAN for beamline component control, a BL-USER-LAN for beamline experimental users and an OA-LAN for the information services. These LANs are interconnected by a firewall system. Since the network traffic and the number of beamlines have increased, we upgraded the backbone of BL-USER-LAN from Fast Ethernet to Gigabit Ethernet. And then, to establish the independency of a beamline and to raise flexibility of every beamline, we also introduced the IEEE802.1Q Virtual LAN (VLAN) technology into the BL-USER-LAN. We discuss here a future plan to build the firewall system with hardware load balancers.

## 1 INTRODUCTION

At the SPring-8 beamline, the network system is an indispensable element. The beamline VME system and workstations communicate with server workstations over the network [1]. The logging data of beamlines are accumulated via the network to the database. Beamline users can control an insertion device and beamline components such as monochromators, mirrors and slits from applications with a TCP/IP socket. Users can transmit experimental data to their home institutions via the network.

The high brilliance X-ray source in SPring-8 made various experiments possible. The large size data were generated as the result of measurement. For example, the data of 256Mbyte/min are generated in the time resolved measurement of 2-dimensional images. The number of users' computers on the BL-USER-LAN has increased because of the rapid construction of beamlines. The network traffic between the BL-USER-LAN and the OA-LAN has also increased. Now 38 beamlines are operational. For the infrastructure of user's experimental environment, the backbone network with high performance was required.

Beamline users connect the computing system and measurement system to the BL-USER-LAN. The IP segments of each beamline were separated. However we could not protect a beamline, e.g.BL01B1, against incorrect packets from other beamline, e.g.BL02B1, if a computer with an IP address of BL01B1 connected with the BL-USER-LAN at BL02B1. We required a BL-USER-LAN with the independency of each beamline.

## 2 ORIGINAL NETWORK

The backbone of the accelerator control network (SR-LAN) is the FDDI [2]. The BL-LAN is separated into four network segments, as A,B,C and D-zone and connected to the FDDI via routers with 10Mbps Ethernet. The backbone of BL-USER-LAN was optical fiber Fast Ethernet. Beamline users connected to the backbone with 10Mbps Ethernet. The OA-LAN is separated into four network segments by a router. A firewall system was introduced to protect the network security at the SR (BL)-LAN and the BL-USER-LAN. The firewall system in the SPring-8 control system consists of five firewalls. One master firewall in the central control room manages other slave firewall modules distributed in the experimental hall. We define the firewall rule that the SR (BL)-LAN is the clean zone, the BL-USER-LAN is the demilitarized zone for experimental users, and the OA-LAN is the public zone for the Internet.

The BL-USER-LAN is separated into 65 IP segments. One is for network management, 62 for all the beamlines, and two for beamline management staff. One C-class IP address is assigned for every beamline. When users' computers access the OA-LAN from the BL-USER-LAN, the firewalls translate the IP address with IP masquerade technique.

## 3 UPGRADE

### 3.1 Gigabit Ethernet

In the winter of 2000, to construct a high performance network system, we upgraded the backbone of BL-USER-LAN from Fast Ethernet to 1Gbps bandwidth optical fiber Gigabit Ethernet. Each beamline uplink connection to the backbone was upgraded from 10Mbps to 100Mbps at the same time.

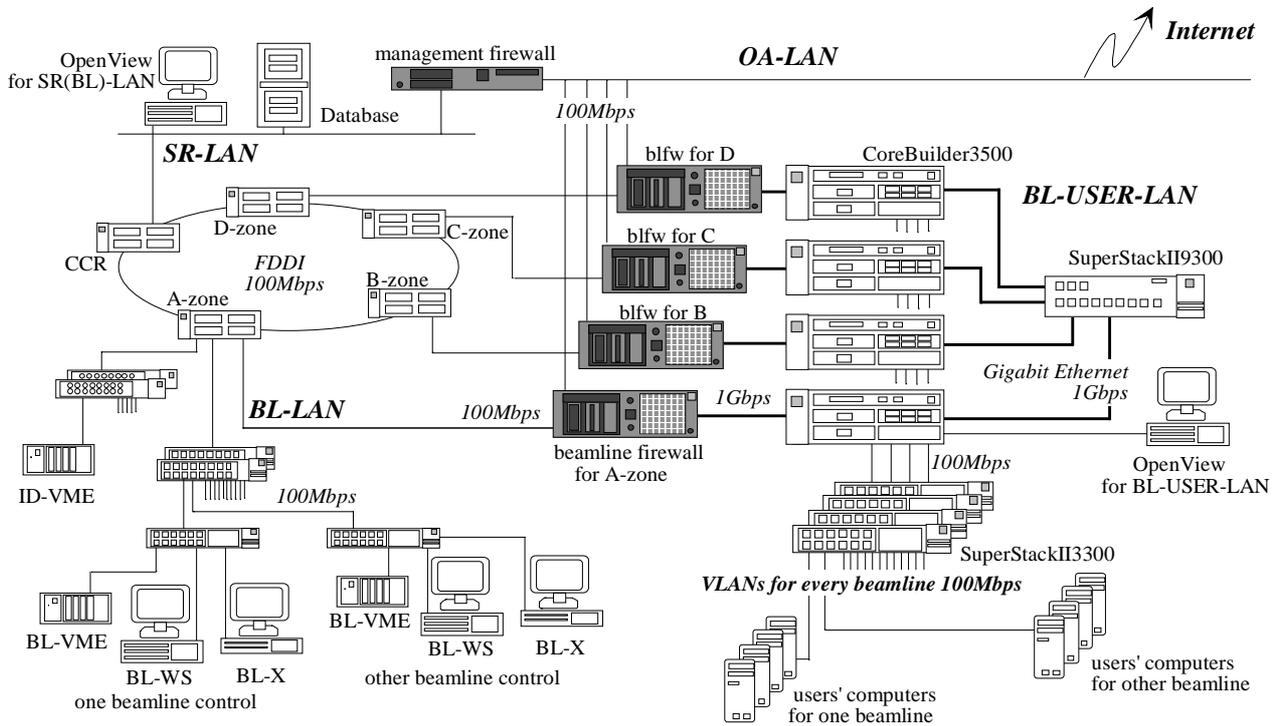

Figure 1: Schematic view of the SPring-8 beamline network

Figure 1 shows the schematic view of the SPring-8 beamline network. There are 500 or more computers and switches in the BL-USER-LAN. Each firewall controls the network traffic for up to 17 beamlines. Firewalls connect to the BL-USER-LAN with 1Gbps bandwidth. In the test bench of Gigabit Ethernet, a firewall throughput was measured to be 170Mbps [3].

From September 2001, the users' group of the laser-electron photon beamline began to transmit experimental data (100Gbyte/day or more) from their experimental system on the BL-USER-LAN to their home institute. We built the network infrastructure responding to the user's request.

### 3.2 VLAN in BL-USER-LAN

When we introduced Gigabit Ethernet, IEEE802.1Q VLAN [4] was applied to the BL-USER-LAN to build a secure and flexible network system.

The VLAN has a mechanism in which each port of a switch is assigned one group, a switch runs like two or more independent switches and each group belongs to an independent broadcast domain. To create many VLANs ranging over many switches, the VLAN-tag is set to the connection port between switches. The field that specifies a VLAN attribute is in an ethernet frame and the VLAN-tag information is set in this field. We can assign the physically separated computers to any VLAN without changing physical cable configuration.

We set a separated IP segment for each beamline with a unique VLAN. No packet from a beamline ever flows to any other beamline since each beamline belongs to an independent VLAN. With the introduction of VLAN, incorrect accesses or broadcast frames from other beamlines are completely shut out.

The introduction of VLAN made flexible network configuration possible. For example, it is easy to set a computing system at the 1km beamline to the same segment in the experimental hall. We only set the same VLAN to the port of switches without changing physical cable configuration.

### 3.3 Network Management

In the BL-USER-LAN, there are four Layer3 switches (3Com [1] CoreBuilder3500), a Gigabit Ethernet switch (3Com SuperStackII9300) and 32 Layer2 10/100Base-TX switches (3Com SuperStackII3300). We manage the BL-USER-LAN using an HP OpenView Network Node

---

[1] http://www.3com.com/

Manager[2] with SNMP in the central control room. We can set easily the VLAN by using Web browser to access the http server built in the switch.

When network trouble arises, OpenView indicates the switch in trouble and the http server shows the port status of the switch. We can find the beamline network trouble from the central control room quickly because each beamline belongs to the unique VLAN.

## 4 PLAN

We recognize that the firewall system is not robust enough and it could be a critical point failure. If a firewall is down in the current network system, users of up to 17 beamlines will have serious damage to their experiment. It will take several hours to recover.

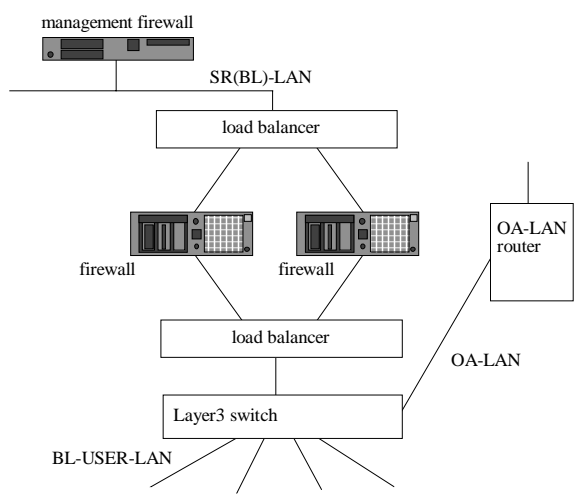

Figure 2: Schematic diagram of the firewall system with hardware load balancers

We are planning to build a redundant firewall system. Figure 2 shows the schematic diagram of the firewall system with hardware load balancers. In this scheme, two firewalls are active and the load balancers distribute packets. Two load balancers check the status of two paths through each firewall. Even if one side path is down with firewall trouble, the network is alive because the load balancer transfers all packets to the effective path.

The layer3 switch in Figure2 connects the SR(BL)-LAN, the BL-USER-LAN and the OA-LAN. The Layer3 switch is required to reject an access from the OA-LAN to the BL-USER-LAN with certain security. The required specifications for the Layer3 switch are as follows:

- We can set up 66 or more VLANs and interfaces.
- 64 or more entries of IP masquerade are possible.
- We can define IP masquerade by the destination of routing.
- The redundant configuration of the management module and the power supply are available.
- Link aggregation between the switches is available.

The access speed from the BL-USER-LAN to the OA-LAN will become higher than the present network system.

## 5 SUMMARY

The Gigabit Ethernet was introduced to meet the increasing network traffic. The VLAN was introduced to establish the independency and flexibility of BL-USER-LAN. The BL-USER-LAN has turned into a system with high speed, secure and flexible.

A future network system will have to support the redundant configuration. We will design the network system including firewall load balancing and high availability.

## REFERENCES

[1] T. Ohata *et al.*, "SPring-8 beamline control system", ICALEPCS'99, Trieste, Italy, 1999, p.666.
[2] T. Fukui *et al.*, "Design and performance of the network system for the storage ring control at SPring-8", ICALEPCS'97, Beijing, China, 1997, p.312.
[3] SPring-8 Annual Report, 1999, p.22.
[4] http://www.ieee802.org/1/

---

[2] http://www.openview.hp.com/